\documentclass[journal,twoside,print]{ieee}

\usepackage{cite}
\usepackage{amsmath}
\usepackage{amssymb}
\usepackage{amsfonts}
\usepackage{algorithmic}
\usepackage{graphicx}
\usepackage{textcomp}
\usepackage{graphicx}
\usepackage{upgreek}
\usepackage{flushend}

\def\mathbi#1{\textbf{\em #1}}
\def\BibTeX{{\rm B\kern-.05em{\sc i\kern-.025em b}\kern-.08em
    T\kern-.1667em\lower.7ex\hbox{E}\kern-.125emX}}
	
\begin{document}

\title{Deep Joint Denoising and Detection for Enhanced Intracellular Particle Analysis}

\author{Y. Yao, I. Smal, \IEEEmembership{Member, IEEE}, I. Grigoriev, A. Akhmanova, and E. Meijering, \IEEEmembership{Fellow, IEEE}
\thanks{This work was supported by the Netherlands Organization for Scientific Research (STW OTP Grant 13391 to E.M. and A.A., NWO Computing Grants 16663 and 17428 to E.M.). The authors gratefully acknowledge NVIDIA Corporation (USA) for their donation of a Titan X Pascal GPU used for the presented research. They also thank SURFsara (the Netherlands) for their support in using the Cartesius GPU computing cluster for the experiments.}
\thanks{Y. Yao was with the Biomedical Imaging Group Rotterdam, Departments of Medical Informatics \& Radiology, Erasmus University Medical Center, Rotterdam, The Netherlands.}
\thanks{I. Smal, I. Grigoriev, and A. Akhmanova are with the Division of Cell Biology, Neurobiology and Biophysics, Department of Biology, Faculty of Science, Utrecht University, Utrecht, The Netherlands.}
\thanks{E. Meijering is with the School of Computer Science and Engineering, University of New South Wales (UNSW), Sydney, Australia (email: erik.meijering@unsw.edu.au).}}

\maketitle

\thispagestyle{empty}

\begin{abstract}
Reliable analysis of intracellular dynamic processes in time-lapse fluorescence microscopy images requires complete and accurate tracking of all small particles in all time frames of the image sequences. A fundamental first step towards this goal is particle detection. Given the small size of the particles, their detection is greatly affected by image noise. Recent studies have shown that applying image denoising as a preprocessing step indeed improves particle detection and their subsequent tracking. Deep learning based particle detection methods have shown superior results compared to traditional detection methods. However, they do not explicitly aim to remove noise from the images to facilitate detection. Thus we hypothesize that their performance could be further improved. In this paper, we propose a new deep neural network, called DENODET (denoising-detection network), which performs image denoising and particle detection simultaneously. We show that integrative denoising and detection yields more accurate detection results. Our method achieves superior results compared to state-of-the-art particle detection methods on the particle tracking challenge dataset and our own real fluorescence microscopy image data.
\end{abstract}


\section{Introduction}\label{sec:introduction}
\IEEEPARstart{T}{ime}-lapse fluorescence microscopy imaging is an essential technique for studying intracellular dynamic processes \cite{Meijering-2008, Liu-2015, Ma-2019}. Current setups of microscopy hardware as well as software make time-lapse image acquisition easier than ever. Despite technological advances, the images still suffer from low signal-to-noise ratio (SNR), due to inevitable quantum phenomena in photon counting and the electronics \cite{Pawley-2006, Meijering-2008}. In addition, the objects of interest are typically small compared to the diffraction limit, their appearance is often very similar, and in various biological scenarios they tend to cluster, making it more difficult to detect them and tell them apart compared to larger objects such as nuclei of migrating cells. These characteristics pose a significant challenge for particle detection and tracking.

Traditionally, particle detection has been done manually, or automatically. The process of manual detection can be time-consuming, prone to errors, and subject to potential human bias. Automatic detection methods have an advantage over manual detection in that they are faster, more reproducible, and potentially more accurate and less biased. Many different automated approaches have been developed to extract particle locations from microscopy images, such as Laplacian-of-Gaussian filtering and thresholding \cite{Basset-2015}, morphological processing \cite{Smal.2008}, wavelet-based processing \cite{Olivo-Marin.2002}, and others, which have been extensively evaluated in various studies \cite{Smal.2010uwt, Ruusuvuori.2010, Stepka-2015, Mabaso.2017}. Recently, deep learning methods have shown superior performance in many bioimage analysis tasks \cite{Xing-2018, Moen-2019, Meijering-2020, Hallou-2021}, including image segmentation, image classification, object detection, and tracking. In particular, convolutional neural networks (CNNs) have been successful in many applications, including particle detection. Several deep learning based particle detection methods have been proposed, such as SpotLearn \cite{Gudla.2017}, Detnet \cite{Wollmann.2019}, and Deep Consensus Network \cite{Wollmann.2021}.

Image denoising is used as a preprocessing step for particle detection and tracking in traditional methods \cite{Luisier2011, Boulanger2010, Smal.2010uwt, Mabaso.2017, Meiniel2018}. Many image denoising methods have been proposed, varying from basic filtering to more sophisticated methods \cite{Mabaso.2017}. A common practical approach to denoise images is to use linear filters, such as the Gaussian. More sophisticated nonlinear methods include median filtering, total-variation based methods, nonlocal filtering, and sparse filtering methods \cite{Meiniel2018}. Current approaches to deep-learning based particle detection do not explicitly exploit denoising. Instead, the raw images are given to the network directly to identify the particle locations. A recent study of image restoration for particle tracking has shown that applying image denoising as a preprocessing step before particle detection improves tracking results, but the nicer-looking denoised images may contain deceiving artifacts, especially with extremely noisy images \cite{Kefer.2021}.

In this paper, we propose a joint denoising and detection method, which uses a single neural network to perform image denoising and particle detection simultaneously. To the best of our knowledge, this is the first work proposing an integrative, end-to-end denoising-detection approach for particle analysis in fluorescence microscopy. Our network is based on a U-Net \cite{Ronneberger.2015} architecture that is extended to a one-encoder-dual-decoders structure exploiting multiple resolutions of the input image. A similar idea was recently used for photoacoustic target localization \cite{Yazdani.2021}, showing state-of-the-art performance on both synthetic and real datasets. Our method is different in several aspects. Since particles have small and less complex appearances, instead of deep encoding by residual blocks, we propose a shallow encoder to extract features from different resolutions. As for the dual decoders, one is for denoising images, and the other is for detecting the particles. We also use skip connections between the encoder and the decoders and explicitly allow knowledge exchange between the two parallel decoders. Compared to a sequential, nonintegrative approach of denoising followed by detection, our method has higher performance and efficiency. 

\section{Background}\label{sec:background}
Deep-learning based detection methods in computer vision are typically developed for natural images \cite{Redmon.2015, Lin.2016, Lin.2017, Tan.2019, Wang.2021}, in which the objects of interest are far larger than particles in microscopy images. State-of-the-art object detection methods can be categorized into two-stage detectors and one-stage detectors. In two-stage methods, approximate object regions are proposed first, which are subsequently used for classification and bounding-box regression to generate object candidates. One-stage methods, on the other hand, predict bounding boxes without an initial region proposal step. They use less computation time but may not be as good at recognizing irregularly shaped objects or clusters of small objects. The bounding boxes for describing the spatial location of objects are commonly taken to be rectangles represented either by the coordinates of their upper-left and lower-right corners or by their center and width and height. Since in particle detection applications the objects of interest are of similar shape and size, it is sufficient to predict the objects centers only.

Newby \cite{Newby.2018} proposed a deep-learning based particle tracking method where particle detection is done by a network comprised of three convolutional layers totaling 6k parameters. The network also uses past and future observations to predict particle locations, which is achieved by a recurrent neural network (RNN) layer. SpotLearn \cite{Gudla.2017} is a method based on the U-Net \cite{Ronneberger.2015} and was developed for spot detection of fluorescence in-situ hybridization (FISH) signals. It uses only three layers in both the encoder and decoder with a total of 402k parameters. The U-Net-like structure is ideal for reconstruction tasks, where the encoder extracts multiple levels of features, and the skip connections between the encoder and decoder help the decoder exploit the learned features in the encoder. Detnet \cite{Wollmann.2019} was adapted from the Deconvolution Network architecture \cite{Noh.2015}, significantly reducing the number of parameters to 17k compared to U-Net with 1.9M parameters, while still achieving high performance. BeadNet \cite{Scherr.2020} was adapted from U-Net to count beads in low-resolution microscopy images. DeepSinse \cite{Danial.2021} is also a simple, multilayer CNN architecture with 21k parameters that detects single molecules. Oktay and Gurses \cite{Oktay.2019} employed a multiple-output CNN (MO-CNN) which outputs the locations and segmentation boundaries of nano-particles in transmission electron microscopy (TEM) images. The Deep Consensus Network \cite{Wollmann.2021} is a one-stage anchor-based detector adapted from RetinaNet \cite{Lin.2017} which utilizes a Feature Pyramid Network \cite{Lin.2016} to extract features at multiple scales. It employs ResNet50 \cite{He.2016} as the backbone network with a squeeze-and-excite mechanism \cite{Hu.2018} for the residuals. Although noise in the images can be a significant problem for particle detection, existing deep-learning methods for particle detection do not explicitly perform image enhancement. However, a recent comprehensive analysis of image restoration for particle tracking showed that improved image quality does lead to higher detection and tracking performance \cite{Kefer.2021}.

\begin{figure*}[!t]
\centering
\includegraphics[width=7.1in]{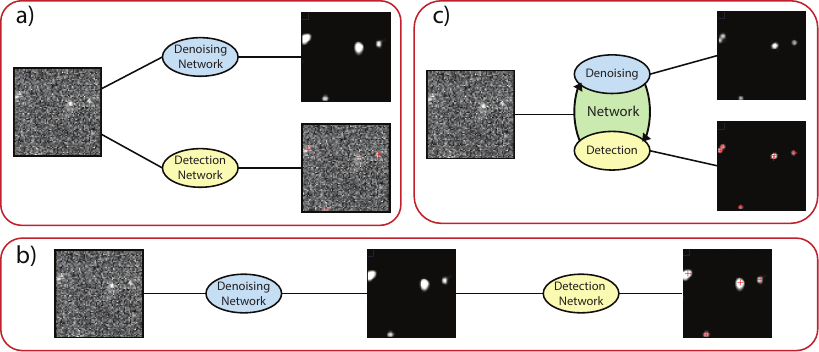}
\caption{Conventional particle detection approaches and our proposed approach. a) Separate denoising and detection networks trained separately. The denoising network takes a noisy input image and outputs an enhanced image. The detection network takes the noisy input image and outputs the particle positions. b) Cascaded approach where the input image is passed to a denoising network and then the enhanced output image is passed to a detection network to extract particle positions. c) Our proposed joint framework for image denoising and particle detection. Our network has an encoder-two-parallel-decoders structure. One of the decoders generates a denoised image while the other detects particles in the noisy input image with skip connections between the encoder and the two parallel decoders.}
\label{fig1}
\end{figure*}

Many deep-learning based image denoising approaches have been developed to enhance natural images and microscopy images recently \cite{Goyal.2020, Dey.2021, Sun.2021}, and have been reviewed by several authors \cite{Goyal.2020, Tian.2020}. Methods such as CARE \cite{Weigert.2018} use high-quality images as ``ground truth'' (GT) together with noisy images to generate data-specific denoising networks. Such methods use supervised training, where the network learns to minimize a distance metric comparing noisy images and their GT targets. However, GT microscopy images are not always available. Thus, synthetic datasets are often used for training these supervised methods. Recently, self-supervised methods have been proposed to train denoising networks without GT images. Noise2Noise (N2N) \cite{Lehtinen.2018} uses pairs of noisy images as inputs to train a denoising network under the assumption that the inputs capture the same content with the only difference being the noise. However, the applicability of this method is limited in practice, as it is difficult to capture the exact same content twice for fast biological processes. Noise2Void (N2V) \cite{Krull.2018} can be trained without the availability of noisy image pairs. Instead, it uses noisy image patches of which the center pixels are masked, and infers the value of these center pixels from their surroundings. This blind-spot network learns to remove only independent noise. However, in practice, many microscopy images additionally contain noisy background structures, causing N2V to produce poor results. A generalized blind-spot network called Struct Noise2Void (SN2V) \cite{Broaddus.2020} uses an extended linear mask to improve the performance and remove locally correlated noise. 

Several deep-learning methods exploit image denoising to help higher-level computer vision tasks. A cascaded network using a joint loss to learn image denoising together with another task (joint training) \cite{Liu.2017} not only boosts the denoising performance but also improves the accuracy of higher-level tasks such as image classification and semantic segmentation. The SDL \cite{Yazdani.2021} method for photoacoustic target localization consists of a shared encoder and two parallel decoders, which outperforms the state-of-the-art on both simulated and real datasets. Liu et al.~\cite{Liu.2021} cascaded denoising and detection for cerebral microbleeds detection in magnetic resonance images (MRIs), using an unsupervised method based on an adaptive squeeze-and-shrink (ASAS) scheme for the denoising task and a modified U-Net for the detection task, showing competitive results. In the approach of Kefer et al. \cite{Kefer.2021}, the noisy image is first processed by a denoising method to get the enhanced image, and the latter is then used to obtain particle locations and tracks. The authors showed that the performance of particle tracking using denoised images improves compared to using noisy inputs and that supervised denoising generally outperforms the self-supervised approach. However, the results also revealed that the qualitatively improved denoised images can produce false-positive particle detections with extremely noisy inputs. A possible alternative, which we propose here, would be to train a network to enhance the image and simultaneously obtain the particle locations. 

As far as we know, existing deep-learning based methods for particle detection are generally trained using noisy images, and do not exploit the noise characteristics to simultaneously perform denoising and particle detections. In other words, the two problems are treated separately (Fig.~\ref{fig1}a). A cascade of two networks (Fig.~\ref{fig1}b) such as joint training \cite{Liu.2017} would be heavily parameterized and cumbersome to train. Thus, in contrast with both the sequential and the cascaded strategy, we propose a novel neural network architecture for particle detection which also simultaneously and explicitly denoises images. Given the input noisy images, the network extracts noise-robust features and outputs the high-SNR and visually faithful denoised images and particle detection results at the same time (Fig.~\ref{fig1}c).
 
\section{Methods}\label{sec:methods}
Reflecting the joint tasks of denoising and detection, our proposed neural network is called DENODET. In the following sections we describe the architecture, the optimization, and the training of the network.

\subsection{Network Architecture}
\begin{figure*}[!t]
\centering
\includegraphics[width=7.1in]{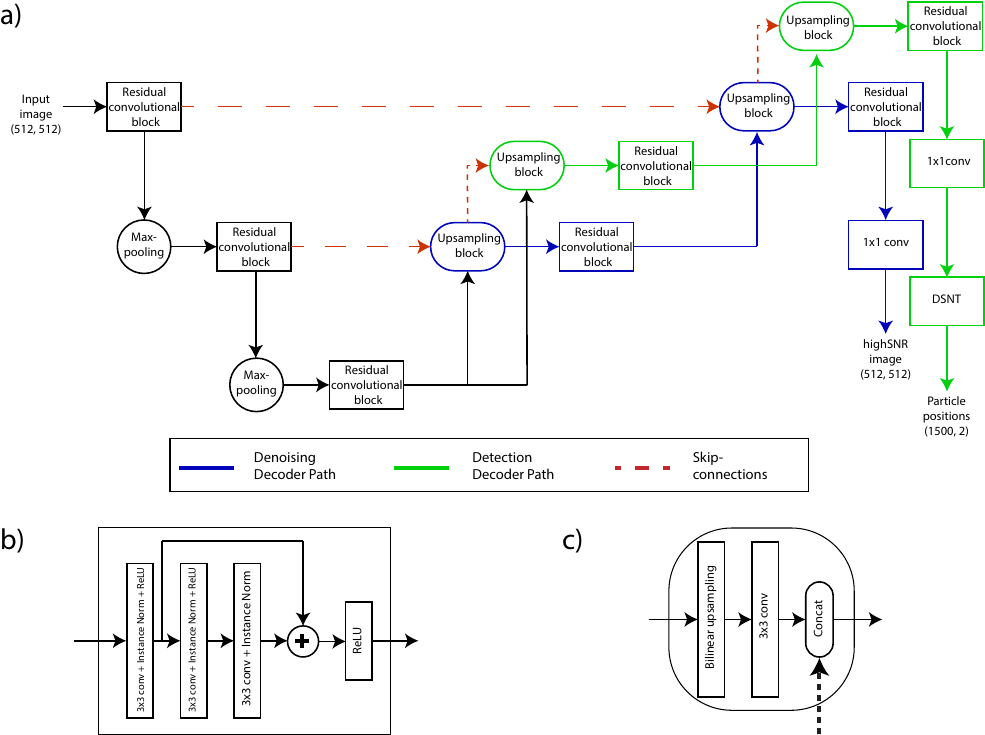}
\caption{Architecture of the proposed network. a) Overview of our one-encoder-dual-decoder network design for performing joint image denoising and particle detection. The solid blue lines indicate the denoising decoder path, the solid green lines indicate the detection decoder path, and the dashed red lines indicate skip connections from the encoder to the two decoders, and from the denoising decoder to the detection decoder in each level. b) Architecture of the residual convolutional block. c) Architecture of the upsampling block.}
\label{fig2}
\end{figure*}

DENODET has a custom, U-shaped architecture consisting of a shared encoder and two parallel decoders (Fig.~\ref{fig2}a). The encoder path extracts multilevel features while successively downsampling the input image, and the two decoder paths use these features to generate an enhanced image and predict particle locations, respectively.

\subsubsection{Shared Encoder}
The encoder architecture uses residual convolutional blocks \cite{Wollmann.2019}. In our network design, a residual convolutional block (Fig.~\ref{fig2}b) consists of three convolutional layers, each using a $3\times3$ kernel, instance normalization, and rectified linear unit (ReLU) activation. The output of the first ReLU is added to the output of the last normalization layer to form a residual and the output is passed to a ReLU activation function. A $2\times2$ max pooling with stride 2 is then applied to downsample the input of the next block. Downsampling enables progressively lower-resolution, higher-scale features to be extracted by the convolution layers.

As the objects of interest in our data are small and have circular shape, with less complicated characteristics than objects in natural or medical images, deep networks such as the original U-Net can easily overfit. Thus, compared to U-Net, we notably decreased the number of feature maps per convolution layer. Specifically, we used 8, 16, and 32 feature maps for the three blocks, respectively. In addition, we constructed a shallow network of three levels, by employing max pooling only two times instead of five times.

\subsubsection{Denoising Decoder}
The denoising decoder (DENO) uses upsampling blocks that increase the size of the extracted low-resolution features to estimate the clean image. An upsampling block consists of a bilinear upsampling layer with a factor of 2, a $3\times3$ convolutional layer, and a concatenation with the corresponding feature map from the encoder. DENO has two levels of upsampling blocks, each followed by a residual convolutional block (Fig.~\ref{fig2}b). Then, a $1\times1$ convolutional layer is applied, which outputs the enhanced image. The number of feature maps in the two levels is 16 and 8, respectively. Each feature map is also passed to the detection decoder of the same level through skip connections, which help refine the final output. This way, a typical encoder-decoder denoising method is developed.

\subsubsection{Detection Decoder}
To also perform particle detection, the network has a parallel decoder which outputs the numerical coordinates of the predicted particle positions. This detection decoder (DET) upsamples the low-dimensional feature maps from the residual convolutional blocks (Fig.~\ref{fig2}a,b), similar to the denoising decoder, except that at each level the output of the denoising layer is also concatenated (Fig.~\ref{fig2}a,c). This way, the concatenation layer in DET incorporates information not only from the encoder, but also from the corresponding denoised feature maps. Similar to DENO, the number of feature maps in the two levels is 16 and 8, respectively. The output of the last residual block is passed to a $1\times1$ convolutional layer which outputs a heatmap representing the likelihood of a particle being present at any position. The numerical coordinates of particle positions are extracted from the heatmap using the differentiable spatial to numerical transform (DSNT) \cite{Nibali.2018}. A DSNT layer adds no trainable parameters and is more suitable for extracting point coordinates than heatmap matching and regression with a fully connected layer \cite{Nibali.2018}.

\subsection{Network Optimization}
The one-encoder-dual-decoder network is jointly optimized via backpropagation \cite{Rumelhart.1995} according to a loss function defined over the output and the reference data. Since there are two parallel paths in DENODET, one from the encoder to DENO which outputs a clean input image, and another from the encoder to DET which outputs a heatmap and particle positions, our network has a separate loss function for each task, which are combined in a joint loss.

\subsubsection{Denoising Loss}
For the denoising decoder, we consider the problem as a pixel-wise classification for which we define the loss term on the output of the denoising decoder as:
\begin{equation}
\mathcal{L}_\text{DENO} = 1-\frac{ 2\mathbi{p}^\text{img} \mathbi{r}^\text{img}+\epsilon} {\mathbi{p}^\text{img} + \mathbi{r}^\text{img}+\epsilon} + \delta\text{BCE}(\mathbi{p}^\text{img}, \mathbi{r}^\text{img}, \beta),
\label{deno_func}
\end{equation}
where the second term is a soft formulation of the standard S{\o}rensen-Dice coefficient between the reference denoised image $\mathbi{r}^\text{img}$ and network prediction $\mathbi{p}^\text{img}$, with a small constant scalar $\epsilon = 10^{-6}$ added to the numerator and denominator to avoid division by zero and to obtain a unit coefficient in the extreme case that both images are entirely zero \cite{Jadon.2020}. This loss performs implicit class balancing compared to the standard cross-entropy loss function and down-weights easy samples compared to that loss. However, training with a Dice loss is not stable especially with small particle objects. Therefore, we use a balanced cross-entropy (BCE) \cite{Xie.2015} loss as a regularizer to help stabilizing the training process, with weight $\delta$.

\subsubsection{Detection Loss}
Our primary task is to predict particle positions as accurately as possible. The loss term defined on the output and reference of the detection decoder is:
\begin{equation}
\mathcal{L}_\text{DET} = ||\mathbi{p}^\text{loc}-\mathbi{r}^\text{loc}||_2 + \lambda\text{JSD}(\mathbi{p}^\text{hm}, \mathbi{r}^\text{hm}_\sigma),
\label{loc_func}
\end{equation}
where $||\cdot||_2$ denotes the Euclidean loss, which directly optimizes the distance between the network predicted positions $\mathbi{p}^\text{loc}$ and actual target positions $\mathbi{r}^\text{loc}$, both of which are in normalized Cartesian coordinates. A Jenson-Shannon divergence (JSD) \cite{Nibali.2018} is added with weight $\lambda$ to act as a distribution regularization on the heatmap between the predicted $\mathbi{p}^\text{hm}$ and a spherical Gaussian distribution $\mathbi{r}^\text{hm}_\sigma$. The Gaussian standard deviation $\sigma$ of the target is set to one pixel as recommended for DSNT \cite{Nibali.2018}.

\subsubsection{Joint Loss}
In our design, the total loss to be minimized combines the two separate losses as follows:
\begin{equation}
\mathcal{L}  = \mathcal{L}_\text{DET}+ \gamma \mathcal{L}_\text{DENO},
\label{tot_func}
\end{equation}
where $\gamma$ is a weight determining the amount of regularization of the detection loss by the denoising loss. This way, the two tasks benefit each other and force the optimization of the encoder parameters to jointly generate a high-SNR image and detect the particles. The denoising decoder helps the encoder to extract more noise-robust features, which in turn enable the detection decoder to output more accurate predictions compared to networks trained directly on the noisy input images. In the experiments, the weights in $\mathcal{L}_\text{DENO}$, $\mathcal{L}_\text{DET}$, and $\mathcal{L}$ were empirically set to, respectively, $\delta = \lambda = 1$ (giving equal unit importance to the regularization terms in the two loss components) and $\gamma = 0.5$ (making the detection loss more prominent than the denoising loss).

\subsection{Network Training}
We trained our model for 100 epochs with a batch size of 4 using the Adam \cite{Kingma.2014} optimizer with an initial learning rate of $10^{-4}$. A learning rate scheduler with a factor of 0.1 and a patience factor of 10 epochs was used to gradually decrease the learning rate during training. Since the loss function has multiple regularization terms that may prevent the training of the network from converging, we initially trained the detection model with noisy images by setting $\gamma=0$, and after about 20 epochs, when it converged to a suboptimal point, we further trained the network end-to-end with the denoising model included.

\section{Experiments}\label{sec:experiments}
The proposed network was evaluated on both synthetic and real images and compared to various state-of-the-art methods. We first describe the datasets, comparison methods, and performance metrics, and then present the results of the experiments on the synthetic and real data.

\subsection{Image Datasets}
To evaluate the network objectively, without requiring excessive manual annotation, we first performed an experiment with synthetic data. To this end we used the particle tracking challenge (PTC) \cite{Chenouard.2014} dataset. More specifically, we used 2D time series (2D+t) representing moving vesicles, receptors, and microtubule plus-ends, as in the PTC. However, the original PTC dataset provides only the noisy images and ground-truth trajectories of the particles, totaling 36 2D+t image sequences (3 particle scenarios, times 4 SNR levels of 1, 2, 4, and 7, times 3 density levels), each having 100 time frames of $512\times512$ pixels. Since our method requires the true noiseless images, we downloaded the Icy plugin called ISBI Challenge Tracking Benchmark Generator, which was used to generate the original PTC dataset, and modified its code to output more images for training including their noiseless versions. For each of the 2D+t scenarios we generated 10 SNR levels, ranging from $\text{SNR}=1$ to $\text{SNR}=10$ in steps of 1, and 5 density levels, with 50, 250, 500, 750, and 1000 particles per image. The resulting synthetic dataset consisted of 150 2D+t image sequences (3 particle scenarios, times 10 SNR levels, times 5 density levels), each having 500 frames of $512\times512$ pixels, making our dataset more than 20 times bigger than the PTC 2D+t dataset. For each particle scenario, we trained our model with the new synthetic data, but used the original PTC data for that scenario for testing.

Since synthetic image data cannot model all spatiotemporal characteristics, imperfections, and variations of time-lapse fluorescence microscopy imaging, we also evaluated performance on real experimental data. Full details on these data are given in our previous work \cite{Yao.2020}. In short, we used HeLa cells expressing fluorescently tagged vesicle markers Rab5, Rab6, Rab11, and the microtubule plus-end marker EB3 \cite{Yang.2017, Martin.2018}, and imaged them using a total internal reflection fluorescence (TIRF) microscope (Nikon, Japan) with a $100\times$ NA 1.49 oil objective lens at 15.4 pixels/$\upmu$m (65 nm/pixel) with exposure time 100 ms at 10 frames/s or with exposure time 500 ms at 2 frames/s. As it is extremely laborious to annotate the positions of all moving particles in these images, we randomly extracted one image frame from each sequence and asked an expert biologist to manually annotate it. In total, we obtained 12 fully annotated images, three for each of the four different markers. We used the model trained on the synthetic data to test the performance on these real images. For the Rab5, Rab6, and Rab11 images, the model trained on the synthetic vesicle data was used, and for the EB3 images, the model trained on the synthetic microtubule plus-end data was used.

\begin{table*}[!t]
\caption{Performance of the compared methods in terms of the F1-score (mean$\pm$standard deviation) for the synthetic dataset. Higher mean values and lower standard deviations imply better performance. Bold indicates best performance per scenario and SNR level. Results of Detnet and DCN are copied from their evaluations on the same data in \cite{Wollmann.2019} and \cite{Wollmann.2021} respectively.}
\resizebox{1\textwidth}{!}{
\begin{tabular}{ccccccccc}
\hline
\textbf{SNR} & \textbf{SOS} & \textbf{Detnet} & \textbf{CARE+Detnet} & \textbf{N2V+Detnet} & \textbf{SN2V+Detnet} & \textbf{DCN} & \textbf{cDENODET} & \textbf{DENODET} \\
\hline
& & & & \textbf{Vesicles} & & & & \\
\hline
1 & 0.261$\pm$0.084 & 0.423$\pm$0.127 & 0.464$\pm$0.110 & 0.440$\pm$0.036 & 0.438$\pm$0.057 & \textbf{0.523$\pm$0.132} & 0.441$\pm$0.063 & 0.444$\pm$0.139 \\
2 & 0.743$\pm$0.051 & 0.939$\pm$0.022 & 0.932$\pm$0.102 & 0.941$\pm$0.031 & 0.943$\pm$0.018 & 0.950$\pm$0.023 & 0.944$\pm$0.034 & \textbf{0.952$\pm$0.018} \\
4 & 0.970$\pm$0.025 & 0.977$\pm$0.016 & 0.981$\pm$0.026 & 0.977$\pm$0.043 & 0.975$\pm$0.017 & 0.993$\pm$0.005 & 0.982$\pm$0.106 & \textbf{0.994$\pm$0.006} \\
7 & 0.966$\pm$0.014 & 0.976$\pm$0.016 & 0.980$\pm$0.016 & 0.978$\pm$0.080 & 0.979$\pm$0.043 & \textbf{0.992$\pm$0.006} & 0.981$\pm$0.025 & 0.989$\pm$0.041 \\
\hline
& & & & \textbf{Receptors} & & & & \\
\hline
1 & 0.173$\pm$0.076 & 0.255$\pm$0.124 & 0.265$\pm$0.240 & 0.233$\pm$0.090 & 0.229$\pm$0.103 & 0.296$\pm$0.105 & 0.437$\pm$0.188 & \textbf{0.440$\pm$0.119} \\
2 & 0.619$\pm$0.116 & 0.802$\pm$0.076 & 0.811$\pm$0.061 & 0.838$\pm$0.113 & 0.840$\pm$0.091 & 0.822$\pm$0.063 & 0.857$\pm$0.078 & \textbf{0.862$\pm$0.084} \\
4 & 0.973$\pm$0.030 & 0.978$\pm$0.017 & 0.980$\pm$0.007 & 0.971$\pm$0.030 & 0.978$\pm$0.074 & 0.993$\pm$0.006 & 0.979$\pm$0.021 & \textbf{0.995$\pm$0.009} \\
7 & 0.984$\pm$0.012 & 0.974$\pm$0.019 & 0.981$\pm$0.012 & 0.977$\pm$0.054 & 0.979$\pm$0.030 & 0.993$\pm$0.006 & 0.979$\pm$0.025 & \textbf{0.993$\pm$0.002} \\
\hline
& & & & \textbf{Microtubules} & & & & \\
\hline
1 & 0.441$\pm$0.116 & 0.481$\pm$0.107 & 0.524$\pm$0.170 & 0.522$\pm$0.072 & 0.530$\pm$0.090 & 0.549$\pm$0.126 & 0.529$\pm$0.172 & \textbf{0.550$\pm$0.080} \\
2 & 0.527$\pm$0.253 & 0.819$\pm$0.035 & 0.834$\pm$0.015 & 0.820$\pm$0.051 & 0.829$\pm$0.074 & 0.829$\pm$0.019 & 0.842$\pm$0.070 & \textbf{0.845$\pm$0.024} \\
4 & 0.751$\pm$0.227 & 0.964$\pm$0.020 & 0.966$\pm$0.067 & 0.960$\pm$0.098 & 0.961$\pm$0.062 & 0.972$\pm$0.018 & 0.973$\pm$0.020 & \textbf{0.977$\pm$0.014} \\
7 & 0.585$\pm$0.334 & 0.977$\pm$0.017 & 0.979$\pm$0.023 & 0.974$\pm$0.400 & 0.970$\pm$0.208 & 0.980$\pm$0.014 & 0.979$\pm$0.033 & \textbf{0.980$\pm$0.013} \\
\hline
\end{tabular}
}
\label{F1}
\end{table*}

\begin{table*}[!t]
\caption{Performance of the compared methods in terms of the RMSE (mean$\pm$standard deviation) for the synthetic dataset. Lower mean values and lower standard deviations imply better performance. Bold indicates best performance per scenario and SNR level. Results of Detnet and DCN are copied from their evaluations on the same data in \cite{Wollmann.2019} and \cite{Wollmann.2021} respectively.}
\resizebox{1\textwidth}{!}{
\begin{tabular}{ccccccccc}
\hline
\textbf{SNR} & \textbf{SOS} & \textbf{Detnet} & \textbf{CARE+Detnet} & \textbf{N2V+Detnet} & \textbf{SN2V+Detnet} & \textbf{DCN} & \textbf{cDENODET} & \textbf{DENODET} \\
\hline
& & & & \textbf{Vesicles} & & & & \\
\hline
1 & 1.962$\pm$0.174 & 1.857$\pm$0.193 & 1.757$\pm$0.138 & 1.732$\pm$0.124 & 1.794$\pm$0.020 & 1.887$\pm$0.121 & 1.680$\pm$0.082 & \textbf{1.620$\pm$0.092} \\
2 & 0.882$\pm$0.108 & 0.766$\pm$0.080 & 0.780$\pm$0.038 & 0.721$\pm$0.092 & 0.700$\pm$0.049 & 0.816$\pm$0.098 & 0.693$\pm$0.070 & \textbf{0.678$\pm$0.101} \\
4 & \textbf{0.393$\pm$0.077} & 0.459$\pm$0.073 & 0.449$\pm$0.032 & 0.460$\pm$0.083 & 0.496$\pm$0.022 & 0.552$\pm$0.043 & 0.421$\pm$0.039 & 0.419$\pm$0.026 \\
7 & \textbf{0.279$\pm$0.194} & 0.427$\pm$0.090 & 0.400$\pm$0.011 & 0.420$\pm$0.037 & 0.417$\pm$0.194 & 0.353$\pm$0.016 & 0.403$\pm$0.028 & 0.410$\pm$0.045 \\
\hline
& & & & \textbf{Receptors} & & & & \\
\hline
1 & 1.891$\pm$0.273 & 1.789$\pm$0.445 & 1.590$\pm$0.118 & 1.800$\pm$0.128 & 1.820$\pm$0.132 & \textbf{1.396$\pm$0.149} & 1.597$\pm$0.221 & 1.587$\pm$0.291 \\
2 & 0.793$\pm$0.132 & 0.693$\pm$0.078 & 0.723$\pm$0.788 & 0.620$\pm$0.082 & 0.618$\pm$0.065 & 0.788$\pm$0.113 & 0.605$\pm$0.063 & \textbf{0.598$\pm$0.079} \\
4 & \textbf{0.320$\pm$0.042} & 0.415$\pm$0.069 & 0.405$\pm$0.164 & 0.477$\pm$0.136 & 0.416$\pm$0.230 & 0.425$\pm$0.016 & 0.412$\pm$0.068 & 0.407$\pm$0.073 \\
7 & \textbf{0.215$\pm$0.054} & 0.440$\pm$0.082 & 0.430$\pm$0.086 & 0.413$\pm$0.076 & 0.408$\pm$0.085 & 0.492$\pm$0.021 & 0.405$\pm$0.069 & 0.402$\pm$0.035 \\
\hline
& & & & \textbf{Microtubules} & & & & \\
\hline
1 & 2.542$\pm$0.149 & 2.419$\pm$0.195 & 2.400$\pm$0.150 & 2.395$\pm$0.140 & 2.351$\pm$0.217 & \textbf{2.185$\pm$0.137} & 2.380$\pm$0.150 & 2.330$\pm$0.123 \\
2 & 2.178$\pm$0.182 & 1.310$\pm$0.150 & 1.173$\pm$0.045 & 1.200$\pm$0.106 & 1.181$\pm$0.282 & 1.354$\pm$0.148 & 1.140$\pm$0.158 & \textbf{1.132$\pm$0.091} \\
4 & 2.038$\pm$0.174 & 0.500$\pm$0.087 & 0.498$\pm$0.092 & 0.610$\pm$0.200 & 0.598$\pm$0.110 & 0.679$\pm$0.085 & 0.475$\pm$0.070 & \textbf{0.469$\pm$0.040} \\
7 & 2.081$\pm$0.221 & 0.411$\pm$0.094 & 0.400$\pm$0.077 & 0.414$\pm$0.095 & 0.420$\pm$0.265 & 0.526$\pm$0.037 & 0.397$\pm$0.090 & \textbf{0.395$\pm$0.088} \\
\hline
\end{tabular}
}
\label{RMSE}
\end{table*}

\subsection{Comparison Methods}
We compared our method to various other methods for particle detection. As a representative example of traditional non-learning methods, we used the SOS detector \cite{Smal.2010uwt, Yao.2017}, which was one of the top performers in the PTC and also performed well in the later single-molecule localization microscopy (SMLM) challenge \cite{Sage.2015,Linde.2019}. It employs wavelet denoising to identify potential spot locations (local maxima) followed by Gaussian PSF fitting in the original (noisy) image to confirm spot existence (based on user-defined intensity thresholds) and obtain the coordinates (the center location of the Gaussian fit). For comparison with other deep-learning based particle detection methods, we included Detnet \cite{Wollmann.2019} and the Deep Consensus Network (DCN) \cite{Wollmann.2021}. Unfortunately, no code or trained models of these networks are publicly available, but their original evaluations in the cited papers were based on the same PTC 2D+t test dataset we are using here as well, allowing us to compare the performance of our methods directly with the published Detnet and DCN results. As neither of these methods use denoising, we also implemented one of them, Detnet, ourselves according to its paper, to be able to test its performance when applied to denoised images. To this end, we trained three state-of-the-art deep-learning denoising methods, namely CARE \cite{Weigert.2018}, N2V \cite{Krull.2018}, and SN2V(3*1) \cite{Broaddus.2020}. Finally, to evaluate the benefits of our parallel approach (Fig.\ \ref{fig1}c) over a cascaded approach (Fig.\ \ref{fig1}b), we also trained a cascaded implementation of our own DENODET, denoted cDENODET, where our denoising network is followed by our detection network. We used the joint training strategy \cite{Liu.2017} to train the cascaded network. Thus, altogether, we compared eight methods, namely SOS, Detnet, CARE+Detnet, N2V+Detnet, SN2V+Detnet, DCN, cDENODET, and the proposed DENODET.

\begin{figure*}[!t]
\centering
\includegraphics[width=\textwidth]{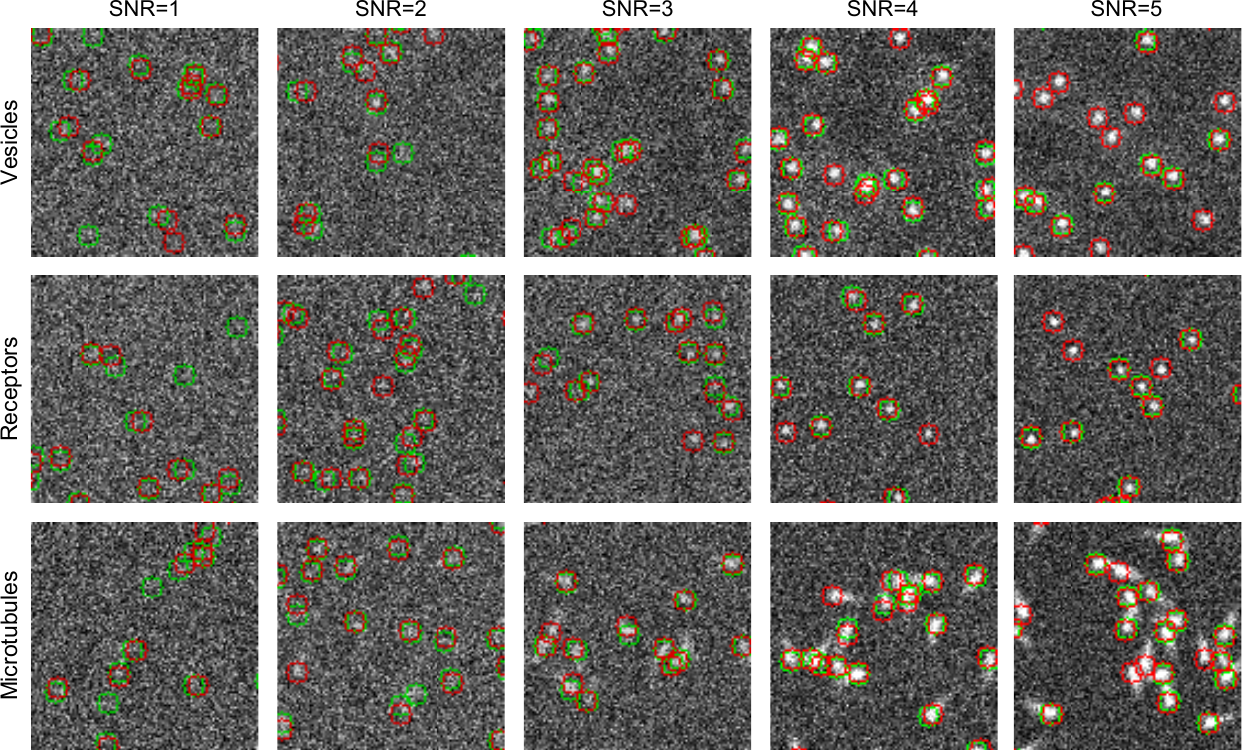}
\caption{Example detection results of DENODET on synthetic images of vesicles, receptors, and microtubule plus-ends, at SNR levels 1, 2, 3, 4, and 5. The red circles indicate the predicted positions and the green circles indicate the reference positions.}
\label{fig3}
\end{figure*}

\subsection{Performance Metrics}
Globally optimal matching of detected particle positions with reference positions (ground-truth positions in the case of synthetic data and annotated gold-standard positions in the case of real data) was performed using the Hungarian algorithm \cite{Kuhn.1955}. The algorithm finds the closest pairs of predicted and reference particles within a distance of 5 pixels \cite{Newby.2018, Wollmann.2019, Wollmann.2021}. This distance is sufficiently small to yield a good matching and is well above the matching error threshold of 1 pixel \cite{Newby.2018}. Predicted particles matched to reference particles within the matching threshold were taken to be true positives (TP), while predicted particles not matched to any reference particle were labeled as false positives (FP), and reference particles not matched to a predicted particle were considered as false negatives (FN). To quantify detection performance, we computed for each image sequence the F1-score:
\begin{equation}
\text{F1} = \frac{2|\text{TP}|}{2|\text{TP}|+|\text{FP}|+|\text{FN}|}.
\label{f1_func}
\end{equation}
We also studied the localization performance in terms of the root mean square error (RMSE) between the matched predicted and reference positions: 
\begin{equation}
\text{RMSE} = \sqrt{\frac{1}{M}\sum^{M}_{i=1}(\mathbi{p}^\text{loc}-\mathbi{r}^\text{loc})^2},
\label{rmse_func}
\end{equation}
where $M$ is the number of matched particles. 

\begin{figure*}[!t]
\centering
\includegraphics[width=0.64\textwidth]{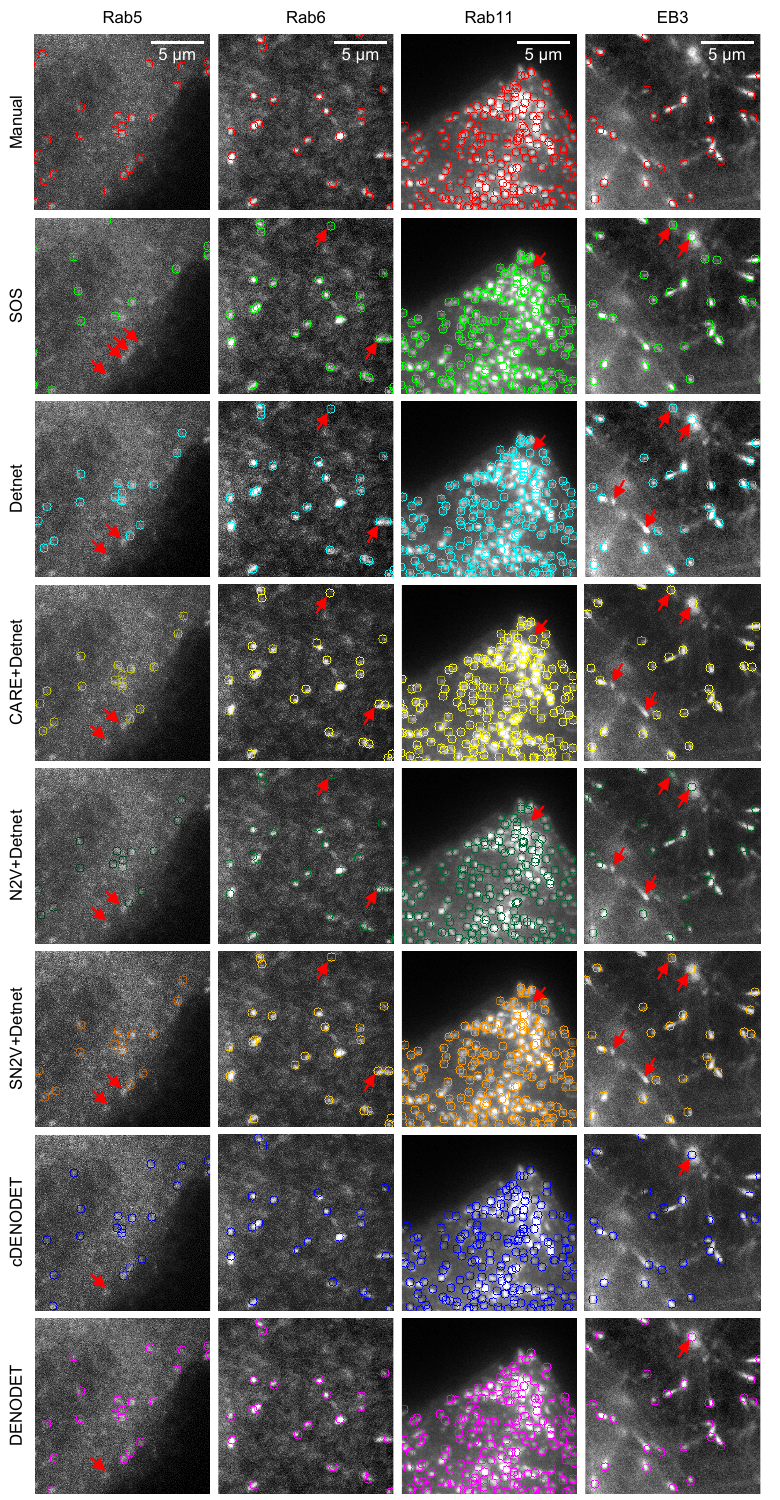}
\caption{Example results of applying detection methods SOS, Detnet, and DENODET to real fluorescence microscopy image data showing different markers Rab5, Rab6, Rab11, and EB3. Reference particle positions manually annotated by an expert biologist are shown in the first row for comparison. Red arrows point at prominent false positive and false negative detections compared to the expert manual annotations.}
\label{fig4}
\end{figure*}

\subsection{Results on Synthetic Data}
We randomly split our synthetic dataset into 70$\%$ for training and 30$\%$ for validation, with stratification to ensure the same split ratio per scenario, SNR, and density level. Testing was done on the original PTC 2D+t dataset. From the results of this experiment, we see that DENODET outperformed the comparison methods in most cases, in terms of both F1 (Table \ref{F1}) and RMSE (Table \ref{RMSE}). Despite the fact that it performs denoising as part of its operation, the traditional SOS method generally showed inferior performance compared to the other tested methods, especially in terms of F1. Also, its performance was lower on the microtubule data than the vesicle and receptor data, which was to be expected, as the Gaussian model used by the method is less suitable for the more elongated microtubule plus-end shapes. This suggests that deep-learning based methods, whether or not they use denoising, are better at learning the right shape model for particle detection than state-of-the-art non-learning based methods.

Our Detnet implementation without denoising showed only slightly decreased performance compared to its original evaluation \cite{Wollmann.2019}, which we assume is due to differences in unpublished details of the method, such as the choice of hyperparameter values, hyperparameter optimization methods, and used software libraries. Application of Detnet to the images denoised by CARE, N2V, and SN2V resulted in improved performance in many cases, though not always. Especially the supervised CARE method seemed to be more beneficial than the unsupervised N2V and SN2V methods. In practice, however, the latter two may be the only viable option when noiseless images are not available for training. DCN generally yielded better results than Detnet, with or without denoising, in terms of detection performance (F1) but not necessarily in localization performance (RMSE), which is in line with observations by the original developers \cite{Wollmann.2021}. The comparison of cDENODET and DENODET, on the other hand, clearly showed the benefits of denoising for our network in terms of both detection and localization. Even in extremely noisy images with SNR as low as 1 or 2, DENODET still does a fair job in pinpointing all particles (Fig.\ \ref{fig3}). However, with all methods, the performance for $\text{SNR}=1$ is substantially lower than for the higher SNR values. This is in agreement with the PTC findings and can be explained from the fact that, even for expert humans, it is very difficult to discern particles in these images, in contrast with images of $\text{SNR}\geq2$.

Additionally, we evaluated the denoising capability of DENODET in comparison with CARE, N2V, and SN2V on the synthetic dataset we generated in-house with the modified PTC image generator (Table \ref{PSNR}). Peek signal-to-noise ratio (PSNR) \cite{Krull.2018} was used to quantify performance. The results suggest that DENODET performs denoising comparable to or better than existing state-of-the-art denoising methods.

\subsection{Results on Real Data}
From the results of the experiment on the real image dataset (Table \ref{real}) we see that DENODET performed comparable to or better than the comparison methods. Here, we compared with SOS, our own implementation of Detnet and cDENODET, as DCN was not available. Denoising with N2V, SN2V, or CARE improved the performance of Detnet. The experiment on real data confirms the observation from the experiment on synthetic data that joint denoising and detection as used in our DENODET method can improve the results (see also Fig.\ \ref{fig4}). DENODET generally performs comparably or better regardless of particle type and density. However, all compared methods start to break down at very low SNR levels (see the Rab5 results in Fig.\ \ref{fig4}).

\begin{table}[!t]
\centering
\renewcommand{\arraystretch}{1.3}
\caption{Performance of the compared methods in terms of PSNR on the synthetic dataset. Higher PSNR values imply better performance. Bold indicates best performance per metric.}
\begin{tabular}{lc}
\hline
\textbf{Method} & \textbf{PSNR}  \\
\hline
CARE      & \textbf{10.274$\pm$2.492}  \\
N2V       & 9.565$\pm$1.724  \\
SN2V      & 9.853$\pm$1.574  \\
DENODET   & 9.897$\pm$1.993  \\
\hline
\end{tabular}
\label{PSNR}
\end{table}

\begin{table}[!t]
\centering
\renewcommand{\arraystretch}{1.3}
\caption{Performance of the compared methods in terms of F1 and RMSE (mean$\pm$standard deviation) on the real microscopy dataset. Higher F1 values and lower RMSE values imply better performance. Bold indicates best performance per metric.}
\begin{tabular}{lcc}
\hline
\textbf{Method} & \textbf{F1}              & \textbf{RMSE}            \\
\hline
SOS              & 0.456$\pm$0.227          & 2.072$\pm$0.313          \\
Detnet           & 0.580$\pm$0.366          & 2.108$\pm$0.407          \\
CARE+Detnet      & 0.619$\pm$0.620          & 1.921$\pm$0.517          \\
N2V+Detnet       & 0.623$\pm$0.201          & 1.864$\pm$0.240          \\
SN2V+Detnet      & 0.633$\pm$0.192          & 1.891$\pm$0.291          \\
cDENONET         & 0.631$\pm$0.052          & 1.910$\pm$0.282          \\
DENODET          & \textbf{0.650$\pm$0.252} & \textbf{1.810$\pm$0.146} \\
\hline
\end{tabular}
\label{real}
\end{table}

\section{Discussion and Conclusion}\label{sec:conclusion}
Deep-learning based methods for particle detection in microscopy images have been used before with promising results, but these methods did not explicitly remove noise from the images, while intuitively one would expect this to improve the detection and localization results. In this paper, we have proposed a new method, DENODET, to test this hypothesis. Our method performs joint denoising (DENO) and detection (DET) by passing information from a single encoder to a dual decoder, one for image denoising and one for particle detection, and also between the two decoders. Unlike other particle detection methods, it uses a fully differentiable and inherent spatial generalization layer, DSNT, to extract particle locations. The denoising path uses the Dice loss with the BCE loss, while the detection path uses the Euclidean loss with JSD, and the joint loss combines the losses of the two paths. We performed a quantitative evaluation of our network on both synthetic and real 2D+t fluorescence microscopy image datasets of various particle types, SNR levels, and densities. For the experiments on synthetic data we modified the image generator of the PTC to obtain a larger training set including noise-free versions of the images and ground-truth particle locations. For the experiments on real data we obtained expert manual annotations for a small subset of images used in previously published studies to serve as gold-standard reference particle locations. Comparisons were made with state-of-the-art non-learning and deep-learning based methods. From the results we conclude that our proposed method yields superior results, confirming our hypothesis that joint denoising and detection is beneficial for particle analysis. We plan to extend our method to be able to also handle 3D+t images, which would require adapting the current 2D CNN, max-pooling, and upsampling layers to 3D layers, and finding ways to make this work without requiring expensive computing hardware. Our proposed approach could potentially also be extended to perform joint denoising and other high-level computer vision tasks, such as cell detection and segmentation, which we aim to explore as well.

\bibliographystyle{IEEEtran}

\begin{thebibliography}{10}

\bibitem{Meijering-2008}
E.~Meijering, I.~Smal, O.~Dzyubachyk, and J.-C. Olivo-Marin, ``Time-lapse
  imaging,'' in \emph{Microscope Image Processing}, Q.~Wu, F.~A. Merchant, and
  K.~R. Castleman, Eds.\hskip 1em plus 0.5em minus 0.4em\relax Academic Press,
  2008, pp. 401--440.

\bibitem{Liu-2015}
Z.~Liu, L.~D. Lavis, and E.~Betzig, ``Imaging live-cell dynamics and structure
  at the single-molecule level,'' \emph{Molecular Cell}, vol.~58, no.~4, pp.
  644--659, 2015.

\bibitem{Ma-2019}
Y.~Ma, X.~Wang, H.~Liu, L.~Wei, and L.~Xiao, ``Recent advances in optical
  microscopic methods for single-particle tracking in biological samples,''
  \emph{Analytical and Bioanalytical Chemistry}, vol. 411, no.~19, pp.
  4445--4463, 2019.

\bibitem{Pawley-2006}
J.~B. Pawley, \emph{Handbook of Biological Confocal Microscopy}, 3rd~ed.\hskip
  1em plus 0.5em minus 0.4em\relax Springer, 2006.

\bibitem{Basset-2015}
A.~Basset, J.~Boulanger, J.~Salamero, P.~Bouthemy, and C.~Kervrann, ``Adaptive
  spot detection with optimal scale selection in fluorescence microscopy
  images,'' \emph{IEEE Transactions on Image Processing}, vol.~24, no.~11, pp.
  4512--4527, 2015.

\bibitem{Smal.2008}
I.~Smal, E.~Meijering, K.~Draegestein, N.~Galjart, I.~Grigoriev, A.~Akhmanova,
  M.~v. Royen, A.~Houtsmuller, and W.~Niessen, ``Multiple object tracking in
  molecular bioimaging by {Rao-Blackwellized} marginal particle filtering,''
  \emph{Medical Image Analysis}, vol.~12, no.~6, pp. 764--777, 2008.

\bibitem{Olivo-Marin.2002}
J.-C. Olivo-Marin, ``Extraction of spots in biological images using multiscale
  products,'' \emph{Pattern Recognition}, vol.~35, no.~9, pp. 1989--1996, 2002.

\bibitem{Smal.2010uwt}
I.~Smal, M.~Loog, W.~Niessen, and E.~Meijering, ``Quantitative comparison of
  spot detection methods in fluorescence microscopy,'' \emph{IEEE Transactions
  on Medical Imaging}, vol.~29, no.~2, pp. 282--301, 2010.

\bibitem{Ruusuvuori.2010}
P.~Ruusuvuori, T.~{\"A}ij{\"o}, S.~Chowdhury, C.~Garmendia-Torres,
  J.~Selinummi, M.~Birbaumer, A.~M. Dudley, L.~Pelkmans, and O.~Yli-Harja,
  ``Evaluation of methods for detection of fluorescence labeled subcellular
  objects in microscope images,'' \emph{BMC Bioinformatics}, vol.~11, no.~1,
  pp. 1--17, 2010.

\bibitem{Stepka-2015}
K.~St{\v{e}}pka, P.~Matula, P.~Matula, S.~W{\"o}rz, K.~Rohr, and M.~Kozubek,
  ``Performance and sensitivity evaluation of {3D} spot detection methods in
  confocal microscopy,'' \emph{Cytometry Part A}, vol.~87, no.~8, pp. 759--772,
  2015.

\bibitem{Mabaso.2017}
M.~A. Mabaso, D.~J. Withey, and B.~Twala, ``Spot detection methods in
  fluorescence microscopy imaging: A review,'' \emph{Image Analysis \&
  Stereology}, vol.~37, no.~3, pp. 173--190, 2017.

\bibitem{Xing-2018}
F.~Xing, Y.~Xie, H.~Su, F.~Liu, and L.~Yang, ``Deep learning in microscopy
  image analysis: A survey,'' \emph{IEEE Transactions on Neural Networks and
  Learning Systems}, vol.~29, no.~10, pp. 4550--4568, 2018.

\bibitem{Moen-2019}
E.~Moen, D.~Bannon, T.~Kudo, W.~Graf, M.~Covert, and D.~Van~Valen, ``Deep
  learning for cellular image analysis,'' \emph{Nature Methods}, vol.~16,
  no.~12, pp. 1233--1246, 2019.

\bibitem{Meijering-2020}
E.~Meijering, ``A bird's-eye view of deep learning in bioimage analysis,''
  \emph{Computational and Structural Biotechnology Journal}, vol.~18, pp.
  2312--2325, 2020.

\bibitem{Hallou-2021}
A.~Hallou, H.~G. Yevick, B.~Dumitrascu, and V.~Uhlmann, ``Deep learning for
  bioimage analysis in developmental biology,'' \emph{Development}, vol. 148,
  no.~18, p. dev199616, 2021.

\bibitem{Gudla.2017}
P.~R. Gudla, K.~Nakayama, G.~Pegoraro, and T.~Misteli, ``{SpotLearn}:
  Convolutional neural network for detection of fluorescence in situ
  hybridization {(FISH)} signals in high-throughput imaging approaches,''
  \emph{Cold Spring Harbor Symposia on Quantitative Biology}, vol.~82, pp.
  57--70, 2017.

\bibitem{Wollmann.2019}
T.~Wollmann, C.~Ritter, J.~N. Dohrke, J.-Y. Lee, R.~Bartenschlager, and
  K.~Rohr, ``Detnet: Deep neural network for particle detection in fluorescence
  microscopy images,'' \emph{Proceedings of the IEEE International Symposium on
  Biomedical Imaging}, pp. 517--520, 2019.

\bibitem{Wollmann.2021}
T.~Wollmann and K.~Rohr, ``Deep consensus network: Aggregating predictions to
  improve object detection in microscopy images,'' \emph{Medical Image
  Analysis}, vol.~70, p. 102019, 2021.

\bibitem{Luisier2011}
F.~Luisier, T.~Blu, and M.~Unser, ``Image denoising in mixed
  {Poisson–Gaussian} noise,'' \emph{IEEE Transactions on Image Processing},
  vol.~20, no.~3, pp. 696--708, 2011.

\bibitem{Boulanger2010}
J.~Boulanger, C.~Kervrann, P.~Bouthemy, P.~Elbau, J.-B. Sibarita, and
  J.~Salamero, ``Patch-based nonlocal functional for denoising fluorescence
  microscopy image sequences,'' \emph{IEEE Transactions on Medical Imaging},
  vol.~29, no.~2, pp. 442--454, 2010.

\bibitem{Meiniel2018}
W.~Meiniel, J.-C. Olivo-Marin, and E.~D. Angelini, ``Denoising of microscopy
  images: A review of the state-of-the-art, and a new sparsity-based method,''
  \emph{IEEE Transactions on Image Processing}, vol.~27, no.~8, pp. 3842--3856,
  2018.

\bibitem{Kefer.2021}
P.~Kefer, F.~Iqbal, M.~Locatelli, J.~Lawrimore, M.~Zhang, K.~Bloom, K.~Bonin,
  P.-A. Vidi, and J.~Liu, ``Performance of deep learning restoration methods
  for the extraction of particle dynamics in noisy microscopy image
  sequences,'' \emph{Molecular Biology of the Cell}, vol.~32, no.~9, pp.
  903--914, 2021.

\bibitem{Ronneberger.2015}
O.~Ronneberger, P.~Fischer, and T.~Brox, ``{U-Net}: Convolutional networks for
  biomedical image segmentation,'' \emph{Proceedings of the International
  Conference on Medical Image Computing and Computer-Assisted Intervention},
  pp. 234--241, 2015.

\bibitem{Yazdani.2021}
A.~Yazdani, S.~Agrawal, K.~Johnstonbaugh, S.-R. Kothapalli, and V.~Monga,
  ``Simultaneous denoising and localization network for photoacoustic target
  localization,'' \emph{IEEE Transactions on Medical Imaging}, vol.~40, no.~9,
  pp. 2367--2379, 2021.

\bibitem{Redmon.2015}
J.~Redmon, S.~Divvala, R.~Girshick, and A.~Farhadi, ``You only look once:
  Unified, real-time object detection,'' \emph{Proceedings of the IEEE
  International Conference on Computer Vision}, pp. 779--788, 2016.

\bibitem{Lin.2016}
T.-Y. Lin, P.~Doll{\'a}r, R.~Girshick, K.~He, B.~Hariharan, and S.~Belongie,
  ``Feature pyramid networks for object detection,'' \emph{Proceedings of the
  IEEE International Conference on Computer Vision}, pp. 2117--2125, 2017.

\bibitem{Lin.2017}
T.-Y. Lin, P.~Goyal, R.~Girshick, K.~He, and P.~Doll{\'a}r, ``Focal loss for
  dense object detection,'' \emph{Proceedings of the IEEE International
  Conference on Computer Vision}, pp. 2980--2988, 2017.

\bibitem{Tan.2019}
M.~Tan, R.~Pang, and Q.~V. Le, ``{EfficientDet}: Scalable and efficient object
  detection,'' \emph{Proceedings of the IEEE International Conference on
  Computer Vision}, pp. 10\,781--10\,790, 2020.

\bibitem{Wang.2021}
C.-Y. Wang, A.~Bochkovskiy, and H.-Y.~M. Liao, ``{Scaled-YOLOv4}: Scaling cross
  stage partial network,'' \emph{Proceedings of the IEEE International
  Conference on Computer Vision}, pp. 13\,029--13\,038, 2021.

\bibitem{Newby.2018}
J.~M. Newby, A.~M. Schaefer, P.~T. Lee, M.~G. Forest, and S.~K. Lai,
  ``Convolutional neural networks automate detection for tracking of
  submicron-scale particles in {2D} and {3D},'' \emph{Proceedings of the
  National Academy of Sciences}, vol. 115, no.~36, pp. 9026--9031, 2018.

\bibitem{Noh.2015}
H.~Noh, S.~Hong, and B.~Han, ``Learning deconvolution network for semantic
  segmentation,'' \emph{Proceedings of the IEEE International Conference on
  Computer Vision}, pp. 1520--1528, 2015.

\bibitem{Scherr.2020}
T.~Scherr, K.~Streule, A.~Bartschat, M.~B{\"o}hland, J.~Stegmaier, M.~Reischl,
  V.~Orian-Rousseau, and R.~Mikut, ``{BeadNet}: Deep learning-based bead
  detection and counting in low-resolution microscopy images,''
  \emph{Bioinformatics}, vol.~36, no.~17, pp. 4668--4670, 2020.

\bibitem{Danial.2021}
J.~S.~H. Danial, R.~Shalaby, K.~Cosentino, M.~M. Mahmoud, F.~Medhat,
  D.~Klenerman, and A.~J.~G. Saez, ``{DeepSinse}: Deep learning-based detection
  of single molecules,'' \emph{Bioinformatics}, vol.~37, no.~21, pp.
  3998--4000, 2021.

\bibitem{Oktay.2019}
A.~B. Oktay and A.~Gurses, ``Automatic detection, localization and segmentation
  of nano-particles with deep learning in microscopy images,'' \emph{Micron},
  vol. 120, pp. 113--119, 2019.

\bibitem{He.2016}
K.~He, X.~Zhang, S.~Ren, and J.~Sun, ``Deep residual learning for image
  recognition,'' \emph{Proceedings of the IEEE Conference on Computer Vision
  and Pattern Recognition}, pp. 770--778, 2016.

\bibitem{Hu.2018}
J.~Hu, L.~Shen, and G.~Sun, ``Squeeze-and-excitation networks,''
  \emph{Proceedings of the IEEE Conference on Computer Vision and Pattern
  Recognition}, pp. 7132--7141, 2018.

\bibitem{Goyal.2020}
B.~Goyal, A.~Dogra, S.~Agrawal, B.~Sohi, and A.~Sharma, ``Image denoising
  review: From classical to state-of-the-art approaches,'' \emph{Information
  Fusion}, vol.~55, pp. 220--244, 2020.

\bibitem{Dey.2021}
S.~Dey, R.~Bhattacharya, F.~Schwenker, and R.~Sarkar, ``Median filter aided
  {CNN} based image denoising: An ensemble aprroach,'' \emph{Algorithms},
  vol.~14, no.~4, p. 109, 2021.

\bibitem{Sun.2021}
H.~Sun, L.~Peng, H.~Zhang, Y.~He, S.~Cao, and L.~Lu, ``Dynamic {PET} image
  denoising using deep image prior combined with regularization by denoising,''
  \emph{IEEE Access}, vol.~9, pp. 52\,378--52\,392, 2021.

\bibitem{Tian.2020}
C.~Tian, L.~Fei, W.~Zheng, Y.~Xu, W.~Zuo, and C.-W. Lin, ``Deep learning on
  image denoising: An overview,'' \emph{Neural Networks}, vol. 131, pp.
  251--275, 2020.

\bibitem{Weigert.2018}
M.~Weigert, U.~Schmidt, T.~Boothe, A.~Müller, A.~Dibrov, A.~Jain, B.~Wilhelm,
  D.~Schmidt, C.~Broaddus, S.~Culley, M.~Rocha-Martins, F.~Segovia-Miranda,
  C.~Norden, R.~Henriques, M.~Zerial, M.~Solimena, J.~Rink, P.~Tomancak,
  L.~Royer, F.~Jug, and E.~W. Myers, ``Content-aware image restoration: Pushing
  the limits of fluorescence microscopy,'' \emph{Nature Methods}, vol.~15,
  no.~12, pp. 1090--1097, 2018.

\bibitem{Lehtinen.2018}
J.~Lehtinen, J.~Munkberg, J.~Hasselgren, S.~Laine, T.~Karras, M.~Aittala, and
  T.~Aila, ``{Noise2Noise}: Learning image restoration without clean data,''
  \emph{Proceedings of the International Conference on Machine Learning}, pp.
  2965--2974, 2018.

\bibitem{Krull.2018}
A.~Krull, T.-O. Buchholz, and F.~Jug, ``{Noise2Void---Learning} denoising from
  single noisy images,'' \emph{Proceedings of the IEEE Conference on Computer
  Vision and Pattern Recognition}, pp. 2129--2137, 2019.

\bibitem{Broaddus.2020}
C.~Broaddus, A.~Krull, M.~Weigert, U.~Schmidt, and G.~Myers, ``Removing
  structured noise with self-supervised blind-spot networks,''
  \emph{Proceedings of the IEEE International Symposium on Biomedical Imaging},
  pp. 159--163, 2020.

\bibitem{Liu.2017}
D.~Liu, B.~Wen, J.~Jiao, X.~Liu, Z.~Wang, and T.~S. Huang, ``Connecting image
  denoising and high-level vision tasks via deep learning,'' \emph{IEEE
  Transactions on Image Processing}, vol.~29, pp. 3695--3706, 2020.

\bibitem{Liu.2021}
H.~Liu, T.~Rashid, J.~Ware, P.~Jensen, T.~Austin, I.~Nasrallah, R.~Bryan,
  S.~Heckbert, and M.~Habes, ``Adaptive squeeze-and-shrink image denoising for
  improving deep detection of cerebral microbleeds,'' in \emph{Proceedings of
  the International Conference on Medical Image Computing and Computer Assisted
  Intervention}, 2021, pp. 265--275.

\bibitem{Nibali.2018}
A.~Nibali, Z.~He, S.~Morgan, and L.~Prendergast, ``Numerical coordinate
  regression with convolutional neural networks,'' \emph{arXiv:1801.07372},
  2018.

\bibitem{Rumelhart.1995}
D.~E. Rumelhart, R.~Durbin, R.~Golden, and Y.~Chauvin, ``Backpropagation: The
  basic theory,'' \emph{Backpropagation: Theory, Architectures and
  Applications}, pp. 1--34, 1995.

\bibitem{Jadon.2020}
S.~Jadon, ``A survey of loss functions for semantic segmentation,'' \emph{Proceedings of the IEEE Conference on Computational Intelligence in Bioinformatics and Computational Biology}, 2020.

\bibitem{Xie.2015}
S.~Xie and Z.~Tu, ``Holistically-nested edge detection,'' \emph{Proceedings of
  the IEEE International Conference on Computer Vision}, pp. 1395--1403, 2015.

\bibitem{Kingma.2014}
I.~K.~M. Jais, A.~R. Ismail, and S.~Q. Nisa, ``Adam optimization algorithm for
  wide and deep neural network,'' \emph{Knowledge Engineering and Data
  Science}, vol.~2, no.~1, pp. 41--46, 2019.

\bibitem{Chenouard.2014}
N.~Chenouard, I.~Smal, F.~{de Chaumont}, M.~Ma\v{s}ka, I.~F. Sbalzarini,
  Y.~Gong, J.~Cardinale, C.~Carthel, S.~Coraluppi, M.~Winter, A.~R. Cohen,
  W.~J. Godinez, K.~Rohr, Y.~Kalaidzidis, L.~Liang, J.~Duncan, H.~Shen, Y.~Xu,
  K.~E.~G. Magnusson, J.~Jald{\'e}n, H.~M. Blau, P.~Paul-Gilloteaux, P.~Roudot,
  C.~Kervrann, F.~Waharte, J.-Y. Tinevez, S.~L. Shorte, J.~Willemse, K.~Celler,
  G.~P. {van Wezel}, H.-W. Dan, Y.-S. Tsai, C.~{Ortiz de Solórzano}, J.-C.
  Olivo-Marin, and E.~Meijering, ``Objective comparison of particle tracking
  methods,'' \emph{Nature Methods}, vol.~11, no.~3, pp. 281--289, 2014.

\bibitem{Yao.2020}
Y.~Yao, I.~Smal, I.~Grigoriev, A.~Akhmanova, and E.~Meijering, ``Deep-learning
  method for data association in particle tracking,'' \emph{Bioinformatics},
  vol.~36, no.~19, pp. 4935--4941, 2020.

\bibitem{Yang.2017}
C.~Yang, J.~Wu, C.~d. Heus, I.~Grigoriev, N.~Liv, Y.~Yao, I.~Smal,
  E.~Meijering, J.~Klumperman, R.~Z. Qi, and A.~Akhmanova, ``{EB1 and EB3}
  regulate microtubule minus end organization and {Golgi} morphology,''
  \emph{Journal of Cell Biology}, vol. 216, no.~10, pp. 3179--3198, 2017.

\bibitem{Martin.2018}
M.~Martin, A.~Veloso, J.~Wu, E.~A. Katrukha, and A.~Akhmanova, ``Control of
  endothelial cell polarity and sprouting angiogenesis by non-centrosomal
  microtubules,'' \emph{eLife}, vol.~7, p. e33864, 2018.

\bibitem{Yao.2017}
Y.~Yao, I.~Smal, I.~Grigoriev, M.~Martin, A.~Akhmanova, and E.~Meijering,
  ``Automated analysis of intracellular dynamic processes,'' \emph{Methods in
  Molecular Biology}, vol. 1563, no.~14, pp. 209--228, 2017.

\bibitem{Sage.2015}
D.~Sage, H.~Kirshner, T.~Pengo, N.~Stuurman, J.~Min, S.~Manley, and M.~Unser, ``Quantitative evaluation of software packages for single-molecule localization microscopy,'' \emph{Nature Methods}, vol.\ 12, no.~8, pp.\ 717--724, 2015.

\bibitem{Linde.2019}
S.~van de Linde, ``Single-molecule localization microscopy analysis with {ImageJ}'', \emph{Journal of Physics D: Applied Physics}, vol.\ 52, no.\ 20, p.\ 203002, 2019.

\bibitem{Kuhn.1955}
H.~W. Kuhn, ``The {Hungarian} method for the assignment problem,'' \emph{Naval
  Research Logistics}, vol.~2, no. 1‐2, pp. 83--97, 1955.

\end{thebibliography}

\end{document}